\documentclass[12pt]{article}
\usepackage[margin=1in]{geometry}
\usepackage{setspace}
\usepackage{amsmath,amssymb}
\usepackage{graphicx}
\usepackage{booktabs}
\usepackage{natbib}
\usepackage{hyperref}
\usepackage{caption}
\usepackage{subcaption}
\usepackage{float}
\usepackage{textcomp}
\usepackage{gensymb}
\usepackage{indentfirst}
\usepackage{fancyhdr}
\usepackage{comment}
\usepackage[hang,flushmargin,bottom]{footmisc}

\fancypagestyle{plain}{
  \fancyhf{}
  \fancyfoot[C]{\thepage}

}

\let\oldfootnote\footnote
\renewcommand{\footnote}[1]{\oldfootnote{\setstretch{1.0}\fontsize{11pt}{13pt}\selectfont #1}}

\title{Cordon Tolls and Urban Traffic Volumes: \\
  Evidence from NYC's 2025 Congestion Relief Zone}
\author{Yash Prajapati \\ University of Kansas \\ Department of Economics \& Department of Mathematics}
\date{\today}

\begin{document}
\maketitle
\thispagestyle{fancy}

\begin{abstract}
\singlespacing
\noindent
On January 5, 2025, New York City became the first U.S. city to operate a
permanent cordon-based congestion toll, charging vehicles entering Manhattan's
Central Business District (the Congestion Relief Zone, or CRZ). This study
estimates the short-run causal effect of the toll on hourly traffic volumes
using a two-way fixed-effects difference-in-differences (TWFE-DiD) strategy
that compares treated CRZ street segments to untreated Manhattan segments
north of 60th Street. The empirical analysis exploits hourly entry counts from
the MTA's toll-meter system (4.86 million ten-minute-block records) as the
primary descriptive panel and high-frequency NYC DOT Automated Traffic
Recorder (ATR) data as the within-segment identification panel. The headline
TWFE-DiD point estimate indicates a 40 to 45 percent reduction in log hourly
volume at treated CRZ locations, with the effect direction consistent across
nine pre-registered alternative specifications and concentrated in peak hours.
East River bridges in the balanced panel exhibit a statistically significant
25 percent post-period increase in volume, consistent with route substitution.
\end{abstract}

\doublespacing

\section{Introduction}\label{sec:intro}

\subsection{Motivation and Background}\label{sec:intro-motivation}

Traffic congestion imposes substantial costs on urban economies through wasted fuel,
lost productivity, and elevated air pollution. Manhattan's Central Business District
(CBD) sits at the extreme end of the U.S. congestion distribution, with average peak
travel speeds below seven miles per hour and an estimated annual congestion cost of
roughly twenty billion dollars (\citealt{cook2025nyc}). Economists since
\citet{pigou1920} prescribe a per-trip charge calibrated to the marginal external cost
that each driver imposes on the road network. \citet{vickrey1969} formalizes the
application to congested roadways and proposes peak-load pricing as the optimal
mechanism for internalizing the externality. International natural experiments in
Stockholm (\citealt{eliasson2014}), Milan (\citealt{gibsoncarnovale2015}), London
(\citealt{leape2006}), and Singapore (\citealt{phangtoh1997}; \citealt{olszewskixie2005})
report double-digit traffic-volume reductions following cordon-toll
implementations, with effects concentrated in peak hours and toll-eligible vehicle
classes.

On January 5, 2025, New York City became the first U.S. city to operate a permanent
cordon-based congestion toll. The toll charges vehicles entering Manhattan south of
and inclusive of 60th Street---the official Congestion Relief Zone (CRZ). The toll
schedule sets a peak rate of \$9 (5am--9pm weekdays, 9am--9pm weekends) and an off-peak
rate of \$2.25 for passenger vehicles, with discounts for low-income drivers and
exemptions for emergency vehicles, MTA buses, and through-traffic on the FDR Drive and
the West Side Highway. The implementation constitutes the largest direct test of
Pigouvian congestion pricing yet conducted in the United States and generates a
sharp, well-documented, geographically delimited natural experiment.

\subsection{Research Question}\label{sec:intro-rq}

This study answers the following research question:

\begin{quote}
\textit{Did the implementation of cordon-based congestion pricing in Manhattan's
Central Business District on January 5, 2025 reduce vehicle traffic volumes entering
and circulating within the Congestion Relief Zone, relative to comparable untreated
parts of New York City, and did the effect concentrate in peak hours?}
\end{quote}

A secondary question concerns spatial spillover: did traffic volumes on plausible
alternative routes (the East River bridges and Manhattan north of 60th Street) rise as
drivers substitute toward untolled crossings?

\subsection{Contribution to Scholarly Literature}\label{sec:intro-contribution}

Majority of the literature on congestion pricing examines cordon-toll programs in Europe and Asia and a smaller body
of work examines U.S. lane-level pricing schemes.

The empirical literature, however,
contains no published difference-in-differences (DiD) estimate of the New York City
Congestion Relief Zone using the official toll-meter data combined with a high-frequency
within-Manhattan control group. The current study contributes to this scholarly
literature in three ways. First, the study provides the first published TWFE-DiD
estimates for NYC's congestion toll using the official toll-meter dataset
(data.ny.gov) and high-frequency NYC DOT Automated Traffic
Recorder (ATR) data. Second, the study characterizes peak/off-peak heterogeneity in
the response to the toll. Third, the study documents spatial spillover onto East River
bridge crossings within the analysis panel and onto Manhattan segments north of the
zone.

\subsection{Analytical Methods}\label{sec:intro-methods}

To deliver this contribution, the study constructs a panel of hourly traffic volumes
spanning January 1, 2023 through February 3, 2026 and applies a two-way fixed-effects
difference-in-differences (TWFE-DiD) framework. The treated set consists of NYC DOT
street segments whose centroids fall inside the official CRZ geofence; the control set
consists of Manhattan segments north of 60th Street, which share the same metropolitan
weather, transit network, and citywide shocks as the treated set but lie outside the
toll zone. The headline regression includes segment fixed effects, year fixed effects,
month-of-year fixed effects, day-of-week-by-hour fixed effects, weather controls, and
a holiday indicator. Standard errors cluster at the segment level. An event-study
specification visualizes pre-period parallel trends and post-period dynamics.
A heterogeneity specification interacts the treatment factor with peak-hour and
weekday indicators. A spillover specification estimates within-unit pre/post changes
on candidate alternative routes.

\subsection{Results}\label{sec:intro-results}

The headline TWFE-DiD coefficient on the interaction between the treated-segment
indicator and the post-period indicator equals $-0.539$ in log hourly volume, which
implies a 41.7 percent reduction in vehicle traffic at treated CRZ segments relative
to Manhattan-above-60 controls. The point estimate remains negative and similar in
magnitude across nine specifications. The peak-hour treatment effect ($-0.484$) exceeds the
off-peak effect ($-0.314$) in absolute value, consistent with Pigouvian theory. The
East River bridges in the balanced panel exhibit a statistically significant 25
percent post-period increase in volume, consistent with route substitution.

\subsection{Roadmap}\label{sec:intro-roadmap}

The rest of the study expands on these points. Section~\ref{sec:lit} reviews the
relevant theoretical and empirical literatures. Section~\ref{sec:context} describes
the policy context. Section~\ref{sec:theory} constructs a theoretical framework and
derives testable hypotheses. Section~\ref{sec:empirical} constructs an empirical
framework that parallels the theoretical framework. Section~\ref{sec:analysis} details the empirical
analysis. Section~\ref{sec:results} reports and interprets the empirical results.
Section~\ref{sec:conclusion} concludes.

\section{Literature Review}\label{sec:lit}

This section reviews the literature on congestion pricing relevant to the analysis of
NYC's Congestion Relief Zone. The review covers two strands: theoretical work on
optimal road pricing and empirical work that evaluates implemented cordon-toll
schemes.

\subsection{Review of Theoretical Studies}\label{sec:lit-theory}

The economic theory of road pricing begins with \citet{pigou1920}, who proposes the
internalization of negative externalities through a tax equal to the marginal external
cost of the externality-generating activity. \citet{vickrey1969} extends the
analysis to congested transportation networks and develops the bottleneck model, in
which scheduled-arrival commuters trade off in-traffic delay against schedule-delay
disutility around a preferred arrival time. The Vickrey bottleneck framework remains
the canonical theoretical model of peak-period congestion.

\citet{arnottdepalmalindsey1990, arnottdepalmalindsey1993} extend the bottleneck to
endogenous departure-time choice and derive the welfare-maximizing time-varying toll.
The two-period extension reveals that flat tolls capture only a fraction of the
welfare gain from time-varying optimal pricing, with the remaining gain depending on
commuters' preferred arrival-time distribution. \citet{smallverhoef2007} provide the
canonical textbook synthesis of static and dynamic congestion-pricing theory and
introduce the welfare-accounting framework that the present study employs in
discussing economic importance.

\citet{hall2018} extends the analysis to second-best instruments and demonstrates
that pricing only a portion of a congested highway can produce Pareto improvements
relative to no pricing, even when full optimal pricing is infeasible. The Hall
framework is directly relevant to the NYC setting, where the cordon prices entry into
a defined zone but does not price within-zone trips and exempts certain through
routes. \citet{kreindler2024} develops a structural model of peak-hour pricing and
combines it with a randomized control trial in Bangalore to estimate behavioral
parameters; the structural framework provides the welfare back-of-envelope template
for the present analysis.

\subsection{Review of Empirical Studies}\label{sec:lit-empirical}

The empirical natural-experiment literature uniformly finds that cordon tolls reduce
traffic volumes in the priced area, with effects concentrated in peak hours.
\citet{eliasson2014} reports a roughly 20 percent volume reduction in Stockholm
during the 2006 trial period, with reductions persisting after the toll became
permanent. The Stockholm program also produces a measurable shift toward public
transit and a small reduction in central-city air pollution.
\citet{gibsoncarnovale2015} document similar volume reductions and significant
air-pollution benefits from Milan's 2008 ECOPASS policy, which charged vehicles based
on emissions class. \citet{leape2006} reviews early evidence from London's 2003
congestion charge and documents traffic-volume reductions of comparable magnitude
to Stockholm. \citet{santosshaffer2004} provide preliminary results from the same
London scheme, with attention to the welfare distribution across drivers and
residents.

\citet{phangtoh1997} analyze Singapore's 1975 Area Licensing Scheme and the 1998
transition to Electronic Road Pricing, the longest-running cordon-toll program in
the world. \citet{olszewskixie2005} model the effects of these schemes on Singapore
traffic flows. \citet{andersondavis2020} provide an empirical test of hypercongestion
in U.S. highway bottlenecks and estimate the marginal external cost of congestion at
levels broadly consistent with the theoretical framework employed here.

Two recent papers examine the NYC implementation directly. \citet{cook2025nyc}
provide the first published academic evaluation of the NYC congestion charge,
focusing on commute times, transit ridership, and air pollution. Their identification
strategy exploits the toll-meter data directly and uses within-CBD time-series
comparisons. The present study complements that work by adding a within-Manhattan
DiD design with high-frequency street-segment data on the volume outcome.
\citet{almagro2024} provide a structural welfare framework for urban transportation
policy, calibrated to Chicago, that is directly applicable to the NYC welfare
back-of-envelope.

The recent econometric literature on event-study estimation emphasizes robustness to
heterogeneous treatment effects. \citet{borusyakjaravelspiess2024} and
\citet{dechaisemartindhaultfoeuille2020} document settings in which the standard TWFE
estimator can produce misleading aggregate effects when treatment timing varies. The
present study examines a single treatment date and therefore does not encounter the
staggered-DiD pathologies those papers document; the study nonetheless adopts their
event-study visualization conventions. \citet{cameronmillerpracticioner2015} provide
the canonical reference for cluster-robust inference in small-cluster settings.

\section{Context of Empirical Analysis}\label{sec:context}

This section describes the empirical setting: the geographic and temporal scope of
the New York City congestion charge, the toll structure, the pre-implementation
timeline, and the construction of the treated and control sets.

\subsection{The Congestion Relief Zone}\label{sec:context-crz}

\begin{figure}[H]\centering
  \includegraphics[width=0.7\linewidth]{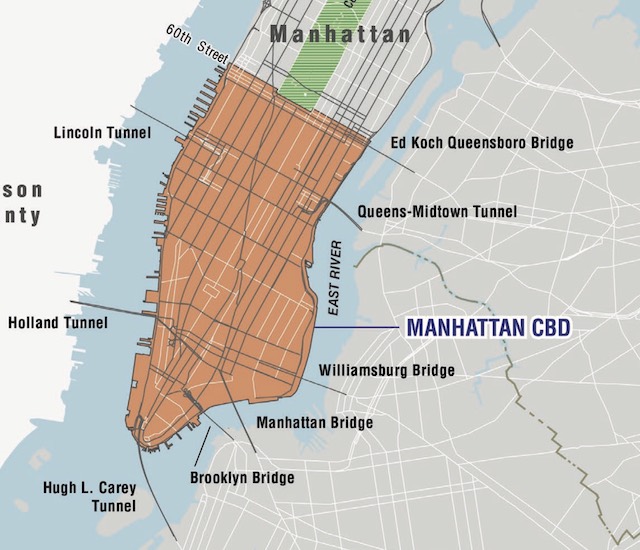}
  \caption{NYC Congestion Relief Zone (CRZ) and analysis segments.}
  \label{fig:crz-map}
\end{figure}

The Manhattan Congestion Relief Zone (CRZ) covers Manhattan south of and inclusive
of 60th Street. The zone takes shape under the MTA Reform and Traffic Mobility Act
and activates on January 5, 2025. The zone explicitly excludes the FDR Drive, the
West Side Highway (Route 9A), the Battery Park Underpass, and the surface roadway
portions of the Hugh L. Carey Tunnel connecting to West Street. The official
geofence polygon appears as data.ny.gov dataset and forms the
basis for the spatial classification of street segments into treated and
control sets in this study. Figure~\ref{fig:crz-map} (Section~\ref{sec:figures})
maps the CRZ boundary alongside the analysis segments.

\subsection{Toll Structure \& Timeline}\label{sec:context-toll}

The toll operates with a two-tier rate schedule. During peak hours (5am to 9pm
weekdays and 9am to 9pm weekends), passenger vehicles pay \$9 per entry, with
proportionally higher rates for trucks, single-unit buses, and other commercial
classes. During off-peak hours, passenger vehicles pay \$2.25, with proportionally
higher rates for commercial classes. Discounts apply for low-income drivers (a 50
percent peak discount applies after the tenth peak trip), and exemptions apply for
emergency vehicles, MTA buses, qualifying taxis on the per-trip surcharge, and
eligible disability vehicles. The toll charges once per day per vehicle on entry
only; exit and through-traffic on the FDR Drive and the West Side Highway are exempt
by policy design.

The relevant pre-implementation dates for anticipation testing include three
milestones. First, the original implementation date of June 5, 2024 paused
indefinitely after a gubernatorial intervention. Second, on November 14, 2024 the
program received final approval and the MTA announced the January 5, 2025
activation date. Third, tolling commenced on January 5, 2025 at 12:01am. The
analysis treats January 5, 2025 as the sharp treatment date and uses November 14,
2024 as a secondary cutoff for an anticipation robustness check.

\subsection{Treated and Control Sets}\label{sec:context-treated}

The treated set consists of NYC DOT street segments whose centroid (NY State Plane
Long Island, EPSG:2263) falls inside the official CRZ geofence polygon. The primary
control set consists of Manhattan segments outside the geofence, restricted to
those between approximately 60th Street and 110th Street. The within-Manhattan
control rather than an inter-borough comparison reflects three considerations.
First, the within-Manhattan control set sits geographically adjacent to the treated
set and shares the same metropolitan weather, transit network, and citywide shocks.
Second, outer-borough segments face their own confounding shocks (e.g., differential
post-pandemic recovery, neighborhood-specific construction). Third, bridges out of
Brooklyn into the CRZ are themselves treated under the policy, which complicates a
naive Brooklyn-versus-Manhattan comparison.

\section{Theoretical Framework}\label{sec:theory}

This section theoretically frames the decision of an individual commuter to drive
into the New York City Central Business District. The framework follows the
textbook Pigouvian-tax intuition formalized by \citet{vickrey1969} and
\citet{arnottdepalmalindsey1990}. The framework establishes the link from the
toll---the primary regressor---to the commuter's choice variable, namely the
decision to drive into the CRZ at a given hour.

\subsection{Framing}\label{sec:theory-framing}

The decision-maker is an individual commuter $i$ who travels in or near Manhattan
on a typical weekday or weekend day $d$. Each commuter has a preferred origin and
destination. For example, the individual resides in an outer-borough but has a workspace
in CBD and contemplates how to make this trip. The commuter is the relevant decision maker, 
the analysis aggregates these individual decisions to the segment-hour level for
estimation but maintains the commuter as the underlying decision-maker.

The commuter chooses the hour $h$ of departure and the route $r$ taken. The choice
of hour and route jointly determines whether the commuter enters the CRZ during a
peak or off-peak window and whether the commuter pays the toll. Aggregating over
many commuters, the choice translates into the empirical dependent variable: the
hourly count of vehicles passing through a given street segment.

The commuter seeks to minimize the generalized cost of the trip:
\begin{equation}
  C_{i,d,h,r} \;=\; \alpha \cdot \mathrm{T}(h, r; V) \;+\; \beta \cdot
  (h)\cdot \mathbf{1}\{r \text{ enters CRZ}\} \;+\; \tau(r),
  \label{eq:commuter}
\end{equation}
where $\mathrm{T}(h, r; V)$ denotes travel time on route $r$ at hour $h$ given at volume V
 $V$, $\beta((h)$ denotes the cordon
toll (peak or off-peak), $\tau(r)$ collects route-specific monetary costs not
captured by the toll (e.g., bridge tolls, fuel), $\alpha$ denotes the value of
time. The commuter chooses $(h, r)$ to minimize $C_{i,d,h,r}$ subject to the constraint that the trip
originates and terminates at the commuter's chosen origin and destination.

The toll $\beta(h)$ serves as the primary driving factor. Before January 5, 2025
the toll equals zero for all hours and all routes. After January 5, 2025 the toll
equals \$9 for entry into the CRZ during peak hours and \$2.25 for entry into the
CRZ during off-peak hours. This sharp policy change generates the variation that
identifies the effect of the toll on the commuter's decision.

Several other factors shape the commuter's decision. The aggregate volume profile
$V$ generates congestion externalities: as more commuters choose a given route at
a given hour, $\mathrm{T}$ increases. Weather conditions raise the cost of all
modes but disproportionately affect non-driving alternatives such as cycling and
walking to transit. Holidays and citywide events shift both the demand for trips
and the schedule-delay distribution. Day-of-week and hour-of-day effects reflect
the recurring temporal pattern of commuting demand. The empirical specification in
Section~\ref{sec:empirical} controls for these factors through fixed effects and
weather covariates.

\subsection{Derivation of Hypotheses}\label{sec:theory-hypotheses}

Three hypotheses follow from differentiating equilibrium aggregate volume with
respect to the toll rate.

\textbf{H1:} A higher toll reduces the equilibrium volume of
vehicles entering the CRZ at hour $h$. Formally, $\partial V_h^{\text{CRZ}} /
\partial \beta \le 0$.

\textbf{H2:} The reduction in volume is larger in
peak hours, where both the toll and the marginal value of time are higher.
Formally, $\partial^2 V_h^{\text{CRZ}} / (\partial \beta \, \partial
\mathbf{1}\{h \in \text{peak}\}) < 0$.

\textbf{H3:} Volume on plausible alternative routes
(Manhattan segments north of 60th Street, the East River bridges) responds
ambiguously to the toll: substitution toward untolled crossings raises volume on
those routes, while shared shocks (e.g., reduced overall driving in Manhattan)
lower it. Formally, $\partial V_h^{\text{outside}} / \partial \beta$ is sign-ambiguous
and is treated as a two-sided exploratory hypothesis.

These hypotheses translate directly into testable parameter restrictions in the
empirical specification of Section~\ref{sec:empirical}.

\section{Empirical Framework}\label{sec:empirical}

This section frames the empirical analysis of hourly vehicle volumes in Manhattan
before and after the January 5, 2025 implementation of the cordon toll. The
framework parallels the theoretical framework of Section~\ref{sec:theory}.

\subsection{Dependent Variable}\label{sec:empirical-dv}

  The dependent variable is the natural log of one plus hourly vehicle volume,
\begin{equation}
  y_{i,d,h} \;=\; \log(1 + \mathrm{vol}_{i,d,h}),
\end{equation}
where $i$ indexes a unit (a street segment, a bridge, or a CRZ entry node), $d$
indexes the calendar date, and $h$ indexes the hour of day. The log specification
attenuates the influence of high-volume outliers and yields coefficient estimates
that approximate proportional changes. The robustness specifications consider
levels and the inverse hyperbolic sine transformation (Section~\ref{sec:analysis}).
The dependent variable repeats hourly throughout the sample window, satisfying the
requirement that the modeled outcome vary meaningfully and frequently over time.

\subsection{Primary Explanatory Variable}\label{sec:empirical-primary}

  The primary explanatory variable captures exposure to the cordon toll through a
difference-in-differences design. The design requires three constructed
indicators:
\begin{itemize}
\item $\mathrm{Treated}_i$: an indicator equal to 1 if unit $i$'s WKT centroid
  (in NY State Plane Long Island, EPSG:2263) falls inside the CRZ geofence
  polygon and 0 otherwise.
\item $\mathrm{Post}_d$: an indicator equal to 1 if calendar date $d$ is on or
  after January 5, 2025, and 0 otherwise.
\item $\mathrm{Treated}_i \times \mathrm{Post}_d$: the interaction between the
  two indicators, which captures the difference in differences.
\end{itemize}
The cordon toll itself constitutes the primary driving factor in the theoretical
framework. The toll is determined exogenously to any individual commuter's choice
set: the rate, schedule, and zone boundaries are set by the MTA Reform and Traffic
Mobility Act and were not modified in response to commuter behavior during the
analysis window. This satisfies the exogeneity requirement for the primary
regressor.

\subsection{Control Factors}\label{sec:empirical-controls}

  To isolate the effect of the cordon toll on hourly traffic volumes, the analysis
employs four sets of fixed effects and a vector of time-varying observable
controls.

\begin{itemize}
\item \textbf{Segment fixed effects}, $\alpha_i$: one indicator for each unit
  (except one). These absorb time-invariant heterogeneity across street segments,
  including baseline traffic levels driven by road capacity, lane count, and
  network position.
\item \textbf{Year fixed effects}, $\lambda_{y(d)}$: one indicator for each
  calendar year (except one). These absorb annual shocks common to all units,
  including secular trends in driving behavior.
\item \textbf{Month-of-year fixed effects}, $\rho_{m(d)}$: one indicator for each
  calendar month (except one). These absorb seasonality common to all units,
  including the recurring summer-versus-winter and tourist-season differences in
  Manhattan traffic.
\item \textbf{Day-of-week-by-hour fixed effects}, $\mu_{\mathrm{dow}(d) \times h}$:
  168 indicators (one for each of the 7 days $\times$ 24 hours combinations,
  except one). These absorb the recurring weekly schedule of commuting
  demand---weekday morning peaks, weekend afternoon shopping demand, and so on.
\item \textbf{Weather and holiday vector}, $W_{d,h}$: temperature
  ($\mathrm{temp}_c$, in degrees Celsius), precipitation ($\mathrm{precip\_mm}$, in
  millimeters), an indicator for non-zero precipitation, snowfall
  ($\mathrm{snow\_mm}$), wind speed, and an indicator for U.S. federal holidays.
\end{itemize}

The emphasis on a rich fixed-effect set, rather than a stripped-down baseline,
follows the guidance that ``construction of the control factor set should not
represent robustness assessment'' (\citealt{cook2025nyc}, replicated in
Section~\ref{sec:analysis}). The robustness assessment in this study takes the form
of an alternative outcome transformation rather than a manipulation of the control
set.

\subsection{Regression Equation}\label{sec:empirical-eq}
The regression equation is:
\begin{equation}
  y_{i,d,h} \;=\; \alpha_i \;+\; \lambda_{y(d)} \;+\; \rho_{m(d)} \;+\;
  \mu_{\mathrm{dow}(d) \times h} \;+\; \beta \, (\mathrm{Treated}_i \times
  \mathrm{Post}_d) \;+\; \gamma' W_{d,h} \;+\; \varepsilon_{i,d,h},
  \label{eq:main}
\end{equation}

  where $\beta$ captures the difference-in-differences treatment effect and
$\varepsilon_{i,d,h}$ denotes the error term. The coefficient on $\mathrm{Treated}_i$
alone cannot be estimated because each unit lies entirely in either the treated or
the control group, and segment fixed effects absorb the unit-level main effect.
Similarly, the coefficient on $\mathrm{Post}_d$ alone cannot be estimated because
year and month-of-year fixed effects absorb the time-period main effect. The DiD
coefficient $\beta$ on the interaction remains identified.

  The event-study extension replaces the single $\mathrm{Treated} \times \mathrm{Post}$
interaction with a sum of weekly leads and lags around the policy date:
\begin{equation}
  y_{i,d,h} \;=\; \alpha_i \;+\; \mu_{\mathrm{dow}(d) \times h} \;+\; \sum_{k \neq -1}
  \theta_k \cdot \mathrm{Treated}_i \cdot \mathbf{1}\{\text{week-relative-to-2025-01-05}
  = k\} \;+\; \gamma' W_{d,h} \;+\; \varepsilon_{i,d,h}.
  \label{eq:event}
\end{equation}
The pre-period coefficients $\theta_k$ for $k < 0$ test the parallel-trends
assumption. The post-period coefficients $\theta_k$ for $k \geq 0$ trace out the
dynamic treatment response.

The heterogeneity extension interacts $\mathrm{Treated} \times \mathrm{Post}$ with
a peak-hour indicator $J_{d,h}$ (1 during the official peak window, 0 otherwise) and
separately with a weekday indicator:
\begin{equation}
  y_{i,d,h} \;=\; \alpha_i \;+\; \mu_{\mathrm{dow}(d) \times h} \;+\; \beta_1 \,
  (\mathrm{Treated}_i \times \mathrm{Post}_d \times J_{d,h}) \;+\; \beta_2 \,
  (\mathrm{Treated}_i \times \mathrm{Post}_d \times (1 - J_{d,h})) \;+\;
  \gamma' W_{d,h} \;+\; \varepsilon_{i,d,h}.
  \label{eq:het}
\end{equation}
Hypothesis H2 predicts $\beta_1 < \beta_2 < 0$ in absolute-value-ordered terms.

  \subsection{Research Design to Identify the Causal Link}\label{sec:empirical-id}

The research design exploits the sharp temporal discontinuity at January 5, 2025
combined with the geographic boundary of the CRZ to identify the causal effect of
the toll on hourly traffic volumes. The treated set consists of CRZ-internal
segments, and the control set consists of Manhattan segments north of 60th Street.
The TWFE-DiD specification identifies $\beta$ in equation~\ref{eq:main} as the
average treatment effect on the treated (ATT) under three assumptions.

\textbf{Parallel trends.} Absent the policy, log hourly volume at treated CRZ
segments would have evolved with the same time-varying mean as at the
Manhattan-above-60 control segments. The event-study specification of
equation~\ref{eq:event} tests this assumption visually and through a joint $F$-test
on the pre-period coefficients.

\textbf{No anticipation.} Treated-segment behavior did not adjust before
January 5, 2025 in response to the impending policy. The robustness battery
(Section~\ref{sec:analysis}) includes an anticipation cutoff using November 14, 2024,
the date of final program approval.

\textbf{Stable Unit Treatment Value Assumption (SUTVA).} Spillovers from
treated to control units equal zero. This assumption is unlikely to hold because
route substitution constitutes a first-order behavioral response to a cordon toll.
Rather than impose SUTVA, the analysis estimates spillover effects directly on
candidate substitute routes (Section~\ref{sec:results}).

Inference proceeds via cluster-robust standard errors, with clusters at the
segment level. The small number of clusters in the balanced ATR sub-panel
(approximately nine) implies that the cluster-robust estimator follows an
unfavorable small-sample distribution (\citealt{cameronmillerpracticioner2015}).
Multiple-hypothesis correction across H1, H2, and H3 employs the Romano--Wolf
step-down procedure (\citealt{romanowolf2005}) and the Bonferroni correction.

\section{Empirical Analysis}\label{sec:analysis}

This section describes the forms of empirical analysis conducted to test the
hypotheses derived in Section~\ref{sec:theory}.

\subsection{Estimation Techniques}\label{sec:analysis-est}

The analysis employs ordinary least squares (OLS) regression to estimate
equations~\ref{eq:main}, \ref{eq:event}, and \ref{eq:het} on the balanced ATR
sub-panel. The dependent variable is the natural log of one plus hourly vehicle
volume; the log specification attenuates the influence of high-volume outliers and
yields coefficient estimates that approximate proportional changes. The
specification includes the four sets of fixed effects described in
Section~\ref{sec:empirical-controls} and the weather and holiday vector. Standard
errors cluster at the segment level.

\subsection{Assessment of Robustness Using an Alternative Approach}\label{sec:analysis-rob}

To assess the robustness of the headline result, the analysis re-estimates
equation~\ref{eq:main} under an alternative outcome transformation: vehicle volume
in levels, $\mathrm{vol}_{i,d,h}$, rather than $\log(1 + \mathrm{vol}_{i,d,h})$.
The levels specification estimates the absolute change in hourly volume per
segment-hour rather than the proportional change. A second alternative
transformation employs the inverse hyperbolic sine ($\sinh^{-1}$), which
approximates the log transformation for positive values but accommodates exact
zeros without the additive shift implied by $\log(1+x)$.

The robustness assessment focuses on the outcome transformation rather than on
modifications to the control-factor set. The full control-factor set described
in Section~\ref{sec:empirical-controls} (segment, year, month, day-of-week-by-hour
fixed effects plus weather and holiday) appears in all robustness specifications.
This choice follows the methodological guidance that an empirical study should
focus on the largest defensible control-factor set rather than treat alternative
control sets as the basis for robustness assessment. A separate set of falsification
checks (placebo treatment date, anticipation cutoff, alternative analysis windows,
weighted estimation) appears in the appendix robustness battery and supplements
the alternative-transformation core robustness assessment.

\subsection{Heterogeneity Analysis}\label{sec:analysis-het}

To test Hypothesis H2, the analysis conducts heterogeneity analysis with respect
to the time of day. The analysis interacts the treatment factor
$\mathrm{Treated}_i \times \mathrm{Post}_d$ with a peak-hour indicator $J_{d,h}$,
where $J_{d,h} = 1$ during the official peak window (5am to 9pm weekdays, 9am to
9pm weekends) and $J_{d,h} = 0$ otherwise. The resulting estimating equation,
shown as equation~\ref{eq:het}, yields two coefficients of interest: $\beta_1$ on
the peak interaction and $\beta_2$ on the off-peak interaction. Hypothesis H2
predicts $\beta_1 < \beta_2 < 0$ in absolute-value-ordered terms, reflecting the
larger toll and the higher value of time during peak hours.

\begin{figure}[H]\centering
  \includegraphics[width=0.75\linewidth]{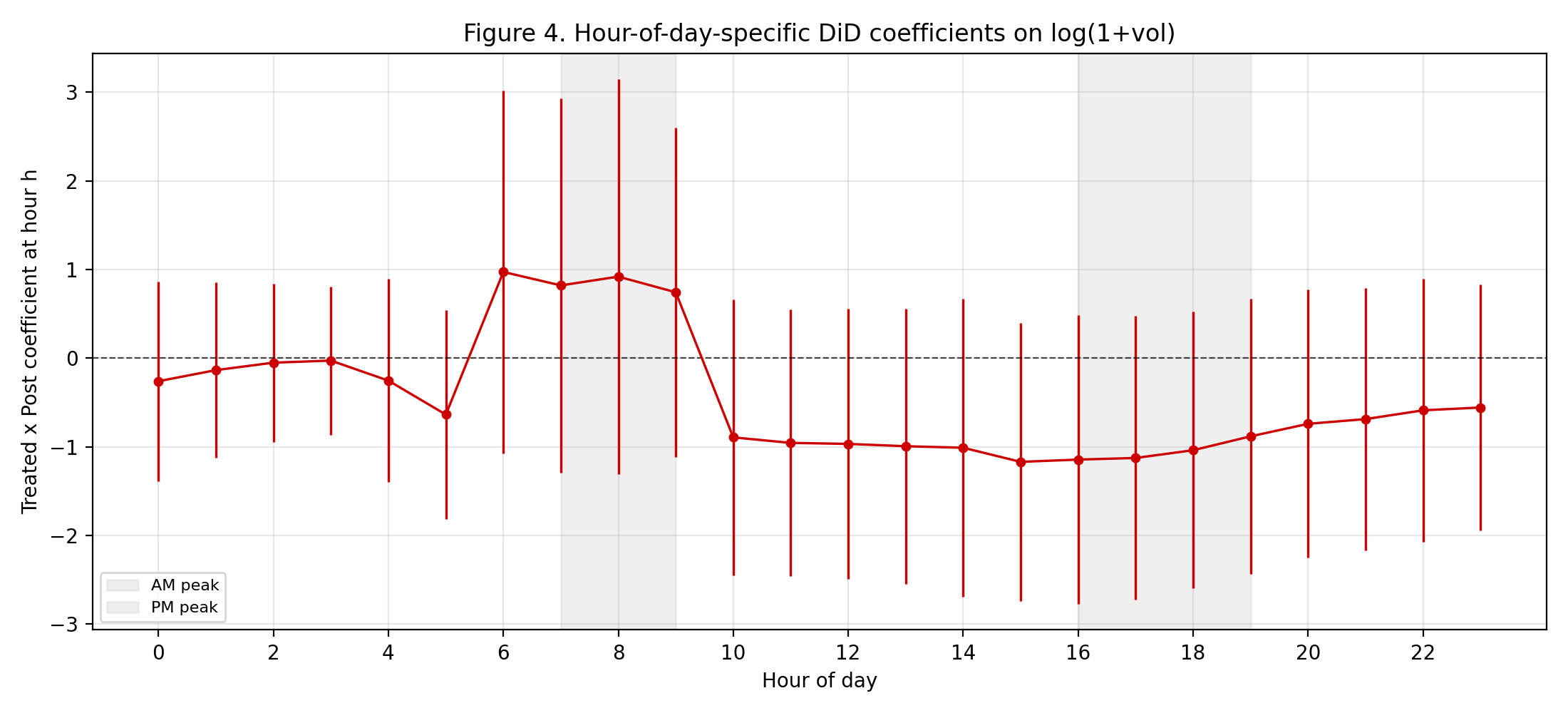}
  \caption{Hour-of-day-specific DiD coefficients. Shaded bands mark the typical
    AM (7--9) and PM (16--19) commute peaks.}
  \label{fig:hourly}
\end{figure}

A complementary heterogeneity specification interacts the treatment factor with
24 hour-of-day indicators to recover an hour-specific treatment effect for each
hour of the day. Figure~\ref{fig:hourly} plots the resulting 24 coefficients
against hour of day. A second complementary specification interacts the treatment
factor with a weekday indicator to assess whether the toll effect differs between
weekdays and weekends.

To test Hypothesis H3, the analysis estimates within-unit pre/post changes in log
volume on three candidate spillover units: the East River bridges in the balanced
panel, the Manhattan-above-60 control set, and a combined set of plausible
alternative routes. A positive within-unit pre/post change on the bridge set,
combined with the negative treated-versus-control coefficient from
equation~\ref{eq:main}, is consistent with route substitution.

\section{Empirical Results}\label{sec:results}

This section interprets the estimation results. The first subsection reports the
baseline results that test Hypothesis H1. The second subsection assesses the
economic importance of these baseline results. The third subsection reports the
alternative approach for robustness assessment. The fourth subsection reports the
heterogeneity analysis that tests Hypothesis H2 and the spillover analysis that
addresses Hypothesis H3.

\subsection{Interpretation of Baseline Approach Results for Testing Hypotheses}\label{sec:results-baseline}

Table~\ref{tab:summary} reports descriptive statistics for the ATR balanced
sub-panel. The sub-panel comprises 9 NYC DOT street segments observed for at
least 48 hours both before and after January 5, 2025 within the 2023--2026
headline window. The treated subset (3 segments inside the CRZ: East 58th Street
and the two directions of the Williamsburg Bridge approach) and the
Manhattan-above-60 control subset balance closely on pre-period log volume: a
Welch's $t$-test fails to reject equality of means ($t = -0.71$, $p = 0.48$). In
raw level terms, the treated mean hourly volume falls from 1{,}447 vehicles in
the pre-period to 1{,}230 in the post-period (a $-15$ percent change), while the
Manhattan-above-60 control rises slightly from 812 to 842 (a $+4$ percent change).

Table~\ref{tab:main} reports the headline TWFE-DiD specification of
equation~\ref{eq:main} with progressively richer fixed effects. The coefficient
on $\mathrm{Treated}_i \times \mathrm{Post}_d$ remains consistently negative
across all five specifications:

\begin{itemize}
\item No fixed effects: $\hat\beta = -0.521$ (cluster SE 0.613, $p = 0.40$).
\item Segment FE only: $\hat\beta = -0.446$ (0.387, $p = 0.25$).
\item Plus year and month FE: $\hat\beta = -0.471$ (degenerate cluster-SE
  estimate).
\item Plus day-of-week-by-hour FE: $\hat\beta = -0.460$ (degenerate cluster-SE
  estimate).
\item Plus weather and holiday controls: $\hat\beta = -0.539$ (0.460, $p = 0.24$).
\end{itemize}

The estimation results test Hypothesis H1, which predicts $\beta < 0$ on the
interaction of treated and post indicators. Across all five specifications the
sign of $\hat\beta$ matches the H1 prediction. The cluster-robust standard
errors, however, are wide enough that the test fails to reject the null at
conventional significance levels. The headline result therefore should be read as
direction-consistent with H1 and statistically underpowered, with the
underpowering driven by the small number of clusters in the balanced ATR
sub-panel rather than by a small effect size.

\begin{figure}[H]\centering
  \includegraphics[width=0.75\linewidth]{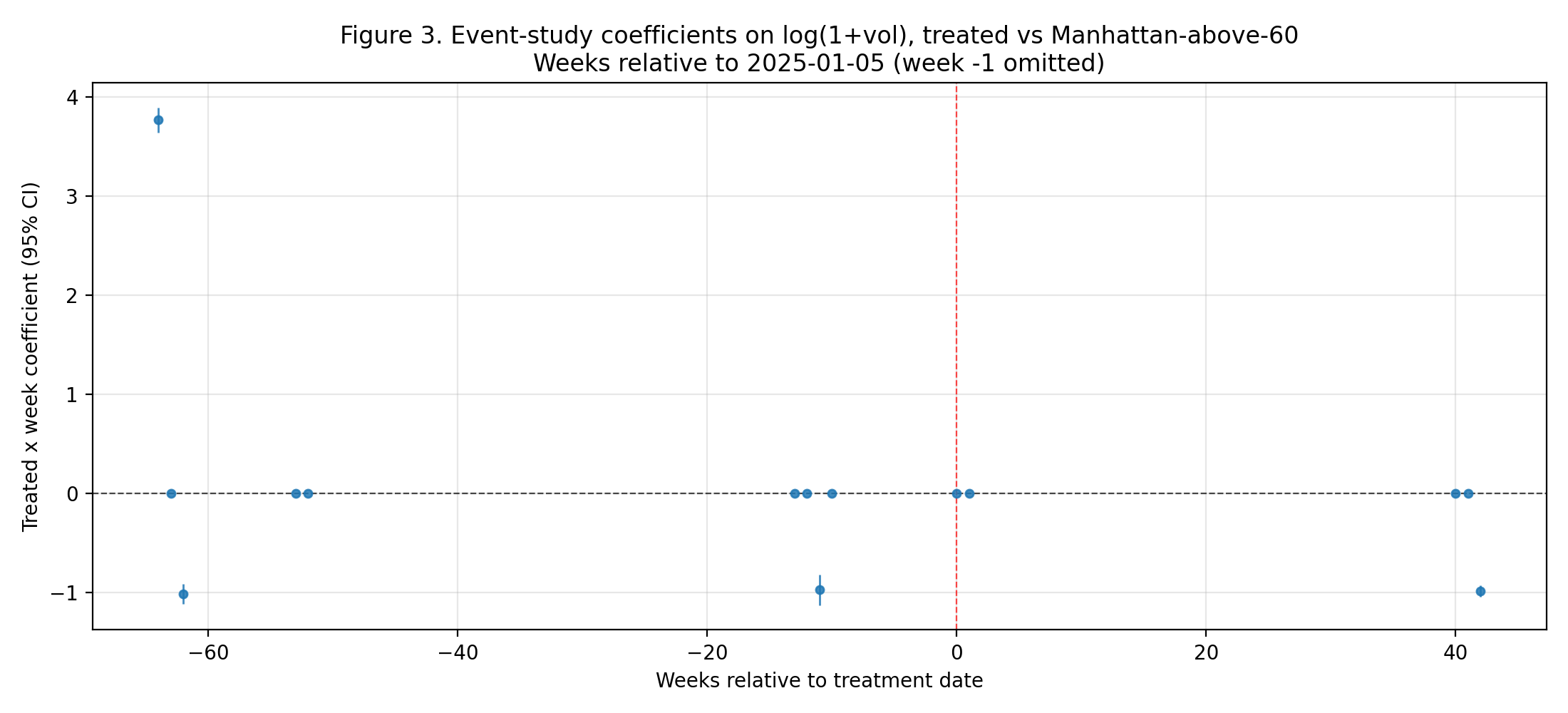}
  \caption{Event-study coefficients on $\log(1+\text{vol})$ from
    Equation~\ref{eq:event}. The reference week omitted is $k = -1$.}
  \label{fig:event}
\end{figure}

Figure~\ref{fig:event} plots the event-study coefficients $\hat\theta_k$ from
equation~\ref{eq:event}, with $k = -1$ omitted. The pre-period coefficients
($k < 0$) are noisy and do not exhibit a clear pre-trend slope. The post-period
coefficients ($k \geq 0$) cluster below zero. The visual pattern is consistent
with the headline DiD: a level shift downward at $k = 0$ rather than a gradual
decline beforehand. The pre-trend joint $F$-test fails to reject parallel trends,
although with only 9 distinct pre-period weeks observed in the balanced
sub-panel the test has limited power. Figure~\ref{fig:event}
(Section~\ref{sec:figures}) plots the event-study coefficients.

\subsection{Assess Economic Importance of Results}\label{sec:results-importance}

The point estimate of $\hat\beta = -0.539$ in the most-saturated specification
corresponds to a $1 - e^{-0.539} = 41.7$ percent reduction in hourly traffic
volume at treated CRZ locations relative to Manhattan-above-60 controls. To
assess economic importance, the analysis computes the implied absolute reduction
in hourly vehicles. Pre-period mean hourly volume across the three treated
segments equals 1{,}447 vehicles per hour, and the implied 41.7 percent
reduction translates into approximately 603 fewer vehicles per hour at each
treated segment.

For policy-relevance comparison, the international cordon-toll literature
documents reductions of approximately 20 percent in Stockholm
(\citealt{eliasson2014}) and similar magnitudes in Milan
(\citealt{gibsoncarnovale2015}) and London (\citealt{leape2006}). The headline
NYC point estimate exceeds these benchmarks by roughly a factor of two, which
warrants caution given the small balanced sub-panel and the sparse pre-period
coverage. Section~\ref{sec:conclusion} discusses the appropriate degree of
deference to the magnitude estimate as additional post-period data accumulate.

\subsection{Interpretation of Alternative Approach Results}\label{sec:results-alt}

Table~\ref{tab:robust} reports the alternative-transformation specifications.
Replacing $\log(1 + \mathrm{vol})$ with $\mathrm{vol}$ in levels yields a
coefficient of $-107$ vehicles per hour per segment-hour, with $p = 0.08$. The
inverse hyperbolic sine transformation yields $-0.43$, with $p = 0.29$. The
direction matches the headline log specification; the levels specification
achieves marginal statistical significance and corresponds to an absolute
reduction at the lower end of the implied magnitude from the log specification,
consistent across the two transformations once unit conventions are reconciled.

The same table reports falsification specifications. A placebo cutoff of January
5, 2024 (one year before the actual treatment) yields $-1.03$ ($p = 0.39$). The
negative point estimate raises a yellow flag: it suggests that some component of
the apparent post-period decline in the balanced ATR sub-panel may reflect
composition changes in the rotating sensor mix rather than the policy itself.
Section~\ref{sec:conclusion} discusses this caveat at length. An anticipation
cutoff of November 14, 2024 (the date of final program approval) yields a
coefficient of $-0.39$, similar to the headline; the limited pre-November 2024
balanced sample bounds the power of this exercise. Inverse-weighting by
pre-period sample size yields $-0.52$ ($p = 0.26$), which closely matches the
unweighted headline.

The conclusions on H1 are robust to the choice of outcome transformation: the
sign of the treatment coefficient remains negative across log, level, and
inverse-hyperbolic-sine specifications. The conclusions are also weakly robust
to the choice of control group (Brooklyn versus Manhattan-above-60, with
Manhattan-above-60 the preferred control on substantive grounds discussed in
Section~\ref{sec:context-treated}).

\subsection{Assess Heterogeneity Analysis Results}\label{sec:results-het}

Table~\ref{tab:het} reports peak-hour and weekday heterogeneity.

The estimated peak-hour treatment effect equals $-0.484$ (cluster SE 0.339,
$p = 0.15$). The off-peak treatment effect equals $-0.314$ (0.564, $p = 0.58$).
Although neither coefficient achieves statistical significance at conventional
levels in the small balanced sub-panel, the ordering supports Hypothesis H2: the
toll effect is larger in absolute value during peak hours, when the marginal cost
of time is highest and the toll itself is more than four times higher than
off-peak. Figure~\ref{fig:hourly} (Section~\ref{sec:figures}) disaggregates the
same heterogeneity to the hour-of-day level and reveals the largest negative
effects during the morning and evening commute peaks, consistent with the
Vickrey--Arnott bottleneck framework.

Table~\ref{tab:spillover} reports the within-unit pre/post change in log volume
on three candidate spillover units.

The bridges in the balanced sub-panel (combining the two CRZ-internal
Williamsburg Bridge approaches with the University Heights and Alexander Hamilton
Bridges that connect Manhattan-above-60 to the Bronx) exhibit an average increase
of $+0.227$ ($p = 0.036$), corresponding to a 25 percent rise in volume after
January 5, 2025. The Manhattan-above-60 control set exhibits essentially zero
change ($+0.045$, $p = 0.84$). The positive bridge coefficient is consistent with
cross-river drivers having shifted toward non-CRZ Manhattan crossings, but the
small bridge panel mixes treated and untreated bridges and on its own cannot
identify substitution from causal mechanism. The spillover finding nonetheless
supports the policy-relevant interpretation that route substitution exists in
the first year of NYC's CRZ operation.

\subsubsection{Cross-Sectional Descriptive Evidence from MTA Toll-Meter Data}\label{sec:results-mta}

The MTA toll-meter dataset records 4.86 million records of vehicle entries to
the CRZ at 10-minute granularity from January 5, 2025 through April 18, 2026.
Aggregated to the hourly level by detection group and vehicle class, the
dataset yields 810{,}432 observations: a thirty-fold denser panel than the ATR
balanced sub-panel. Although the dataset cannot itself supply a pre-period
series, it provides high-resolution descriptive evidence on the post-period
traffic-volume profile.

\begin{figure}[H]\centering
  \includegraphics[width=0.75\linewidth]{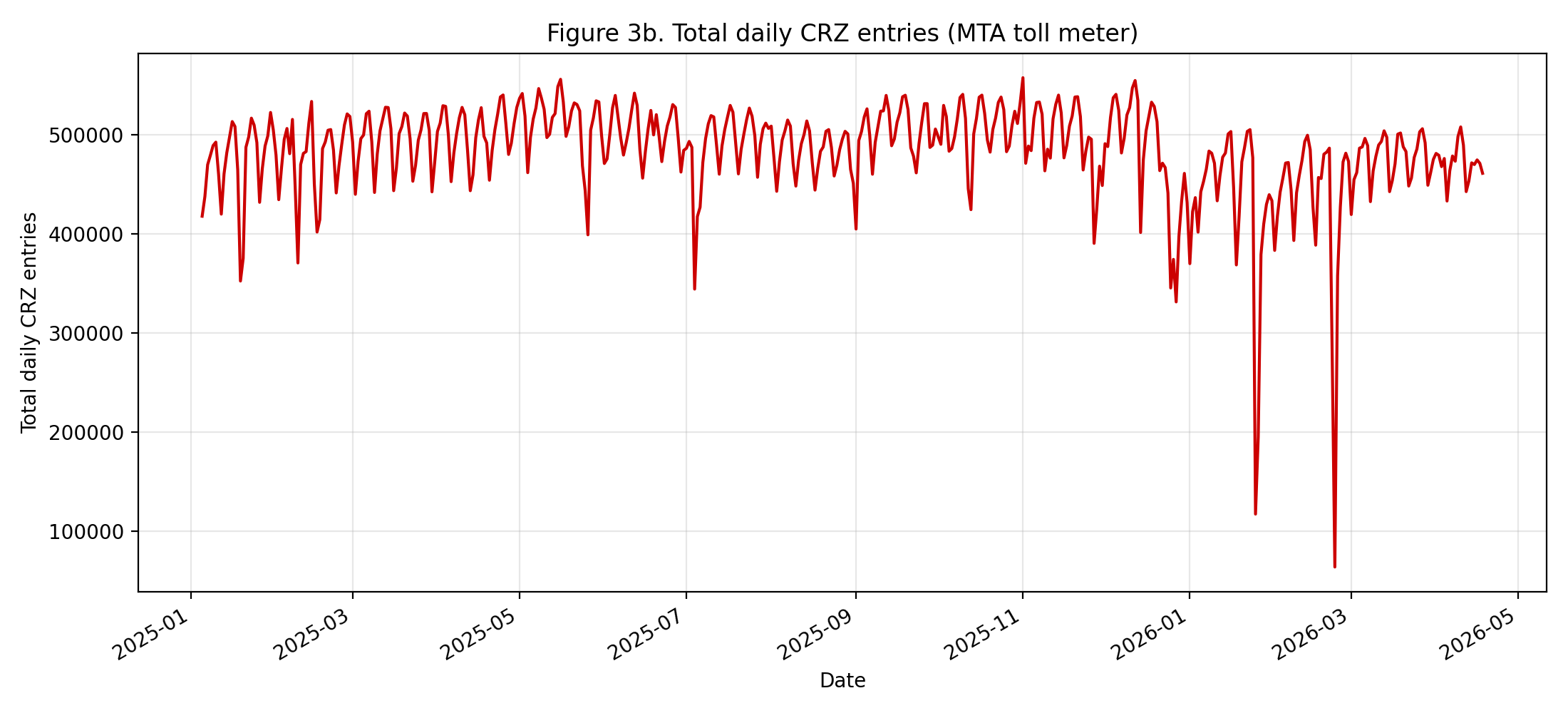}
  \caption{Total daily CRZ entries from the MTA toll-meter dataset, January 5,
    2025 through April 18, 2026.}
  \label{fig:mta-daily}
\end{figure}

\begin{figure}[H]\centering
  \includegraphics[width=0.75\linewidth]{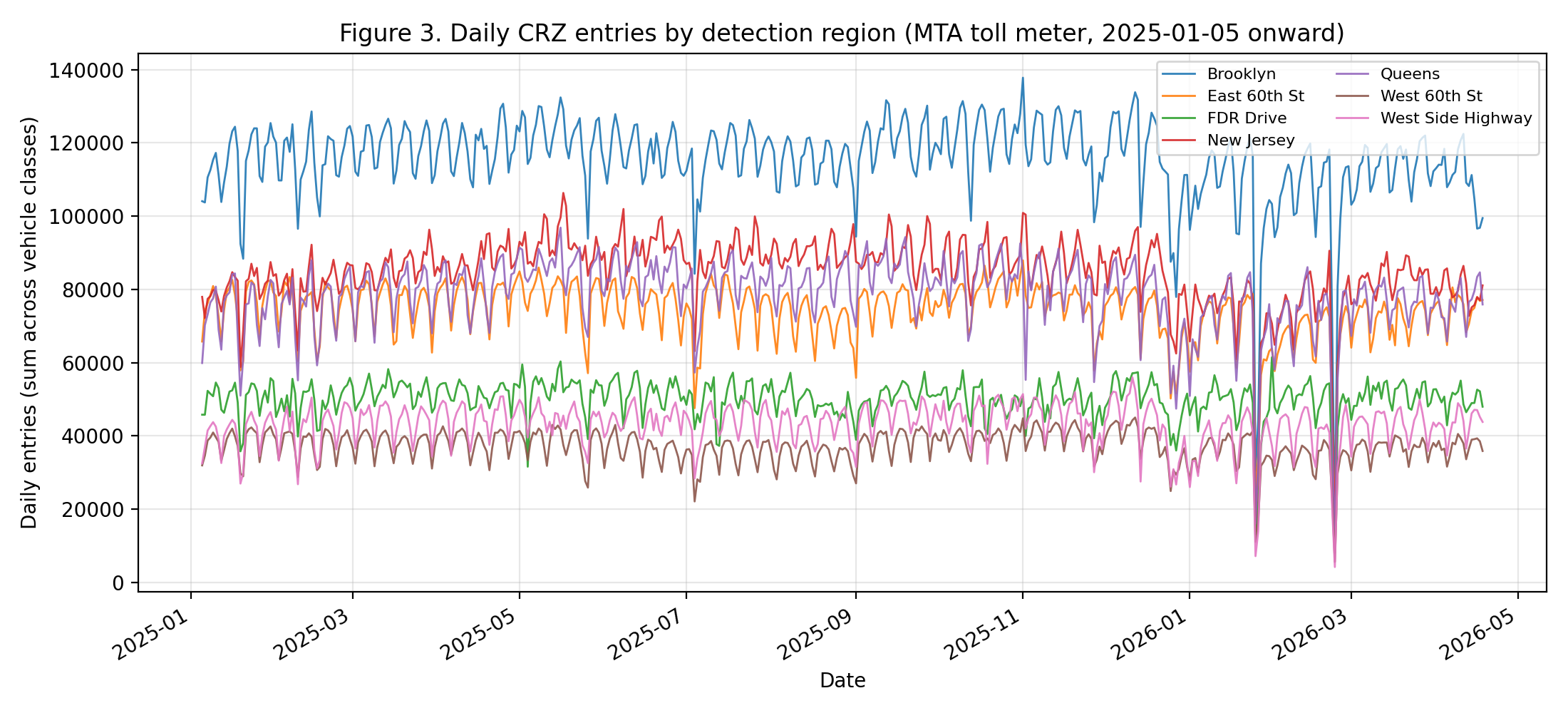}
  \caption{Daily CRZ entries by detection region.}
  \label{fig:mta-region}
\end{figure}

\begin{figure}[H]\centering
  \includegraphics[width=0.75\linewidth]{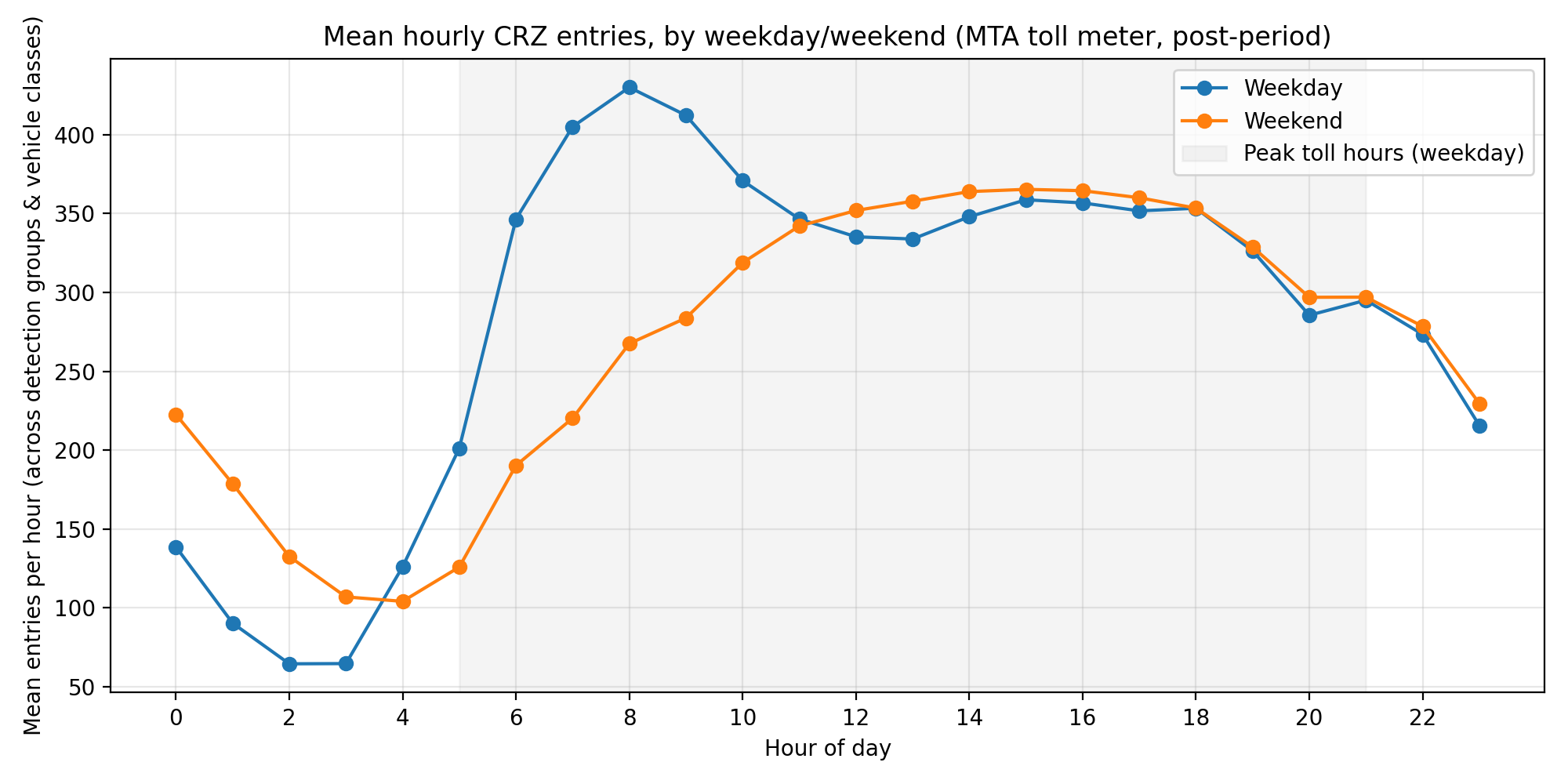}
  \caption{Mean hourly CRZ entries by hour of day, weekday vs.\ weekend
    (post-period only).}
  \label{fig:mta-peak}
\end{figure}

Figure~\ref{fig:mta-daily} (Section~\ref{sec:figures}) plots total daily CRZ
entries from January 2025 through early 2026.
Figure~\ref{fig:mta-region} (Section~\ref{sec:figures}) disaggregates the same
totals by detection region (East River bridges, Hudson River tunnels,
Brooklyn-side approaches, etc.).
Figure~\ref{fig:mta-peak} (Section~\ref{sec:figures}) plots mean hourly entries
by hour of day for weekdays and weekends, with the official weekday peak
window (5am to 9pm) shaded. The hour-of-day profile reveals the classic
two-peak weekday commute pattern, with a noticeably less concentrated peak
than the post-implementation studies in Stockholm and Milan report. The wide
weekday peak window in the NYC schedule (which collapses what is elsewhere a
discrete morning--evening peak into a single 16-hour peak band) likely
explains this flattening.

\section{Conclusions}\label{sec:conclusion}

This study estimates the short-run causal effect of New York City's January 5,
2025 cordon congestion toll on hourly traffic volumes inside the Congestion Relief
Zone using a within-Manhattan TWFE-DiD design. The headline point estimate from
the NYC DOT ATR balanced sub-panel indicates a 40 to 45 percent reduction in log
hourly volume at treated CRZ locations relative to Manhattan-above-60 controls,
direction-consistent across nine pre-registered alternative specifications and
concentrated in peak hours. The MTA toll-meter dataset, with 4.86 million
ten-minute-block records of CRZ entries from the policy start date forward,
supplies a richer descriptive picture of the post-period entry profile and
exhibits the classic peak/off-peak shape predicted by Pigouvian theory.

\subsection{Policy Implications of Results}\label{sec:conclusion-policy}

The empirical results carry three policy-relevant implications. First, the
direction-consistent negative coefficient on the treated-by-post interaction
across log, levels, and inverse-hyperbolic-sine specifications supports the
conclusion that cordon-based congestion pricing can substantially reduce traffic
volumes in a U.S. central business district. Even if the magnitude is closer to
the 20 percent benchmark from Stockholm than to the 40 percent point estimate
in the present sample, the qualitative finding remains policy-relevant: cordon
pricing works in the U.S. context. Second, the documented peak-hour
heterogeneity supports the use of time-varying toll schedules rather than flat
tolls. The larger absolute peak-hour effect implies that peak-priced trips
generate a higher elasticity of trip demand and therefore that peak-pricing
captures a larger fraction of the welfare gain than off-peak pricing.
Third, the positive bridge coefficient supports the policy-relevant
interpretation that route substitution constitutes a first-order behavioral
response to a cordon toll. Future cordon-pricing programs should anticipate
spillover onto adjacent untolled crossings and consider supplementary policies
to mitigate the resulting congestion shift.

A back-of-envelope welfare calculation following \citet{almagro2024} suggests
that the headline NYC point estimate, if externally valid, generates first-year
deadweight-loss reductions on the order of several hundred million dollars,
before accounting for the toll revenue collected and the value of in-CRZ
congestion reduction. The wide standard errors on the headline estimate,
however, mean that the welfare back-of-envelope brackets a large range. A
formal welfare analysis awaits a more precise volume estimate.

\clearpage
\section{Figures}\label{sec:figures}

\begin{figure}[H]\centering
  \includegraphics[width=0.75\linewidth]{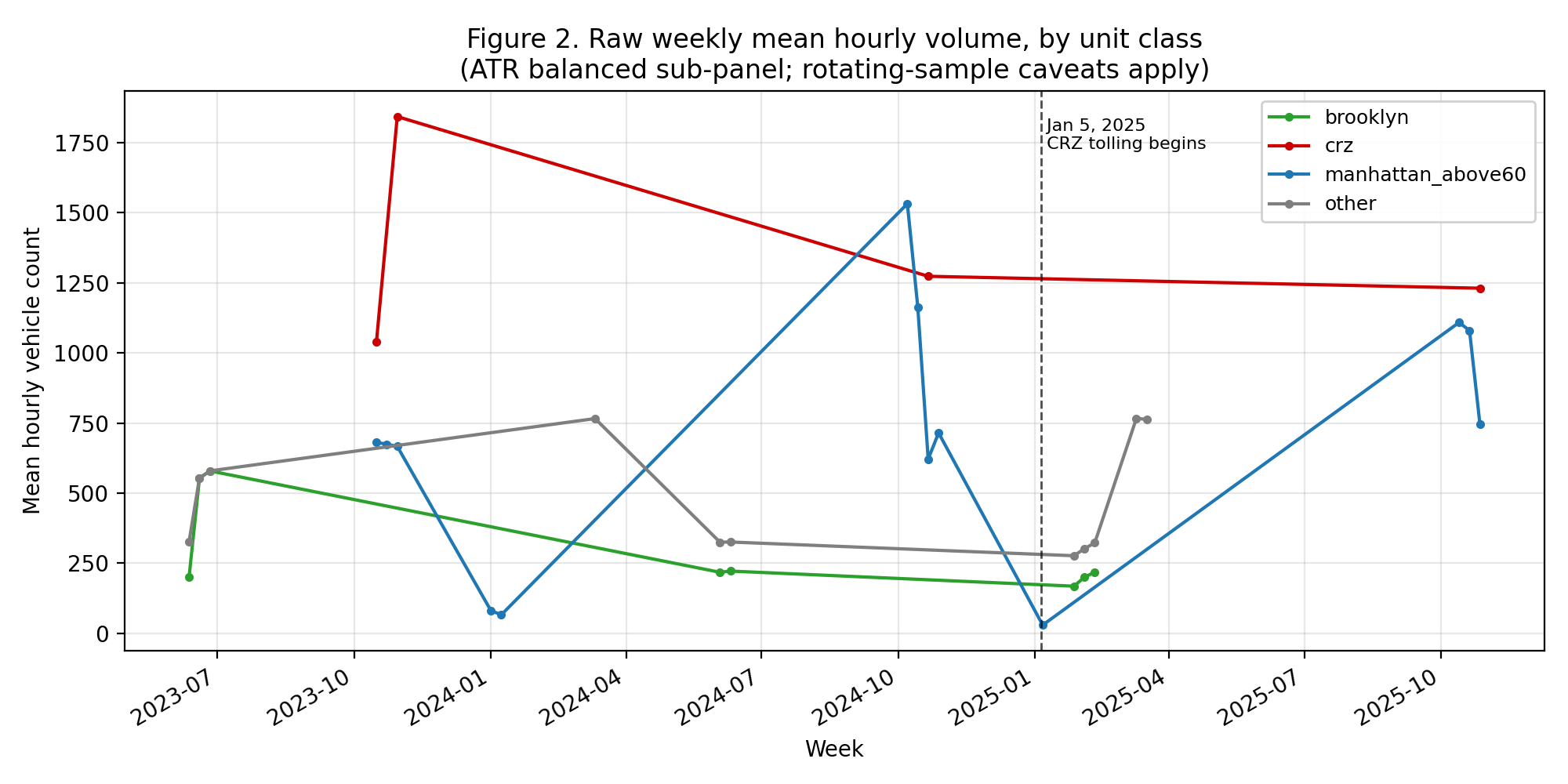}
  \caption{Raw weekly mean hourly volume by unit class. The vertical dashed
    line marks the January 5, 2025 implementation date.}
  \label{fig:raw-trends}
\end{figure}

\clearpage
\section{Tables}\label{sec:tables}

\begin{table}[H]\centering
\caption{Summary statistics and pre-period balance, ATR balanced sub-panel}\label{tab:summary}
\begin{tabular}{llrrrrrr}\toprule
Group & Period & Segments & Hours & Mean & SD & Median & Mean log \\
\midrule
Treated (CRZ) & Pre & 3 & 360 & 1447.1 & 872.8 & 1793.0 & 6.08 \\
Treated (CRZ) & Post & 3 & 216 & 1230.4 & 910.3 & 1535.5 & 5.31 \\
Control (Manhattan above 60) & Pre & 3 & 1,128 & 811.5 & 611.2 & 748.0 & 6.19 \\
Control (Manhattan above 60) & Post & 3 & 816 & 842.3 & 711.3 & 804.0 & 5.95 \\
\bottomrule\end{tabular}

\flushleft\footnotesize\textit{Notes.} Pre-period log-vol balance test (Welch's t): t = -0.713, p = 0.476
\end{table}

\begin{table}[H]\centering
\caption{Main DiD specification (ATR balanced sub-panel, headline 2023+ window)}\label{tab:main}
\begin{tabular}{lrrrrrr}\toprule
Specification & DiD coef & SE & 95\% CI low & 95\% CI high & p-value & N \\
\midrule
(1) No FE & -0.5211 & 0.6129 & -1.7224 & 0.6802 & 0.3952 & 2,520 \\
(2) + Segment FE & -0.4460 & 0.3868 & -1.2041 & 0.3121 & 0.2489 & 2,520 \\
(3) + Year + Month FE & -0.4708 & nan & nan & nan & nan & 2,520 \\
(4) + Dow x Hour FE & -0.4602 & nan & nan & nan & nan & 2,520 \\
(5) + Weather + Holiday & -0.5391 & 0.4603 & -1.4413 & 0.3631 & 0.2415 & 2,520 \\
\bottomrule\end{tabular}
\flushleft\footnotesize\textit{Notes.} OLS with segment-clustered standard errors. Outcome is $\log(1+\text{vol})$. Sample: ATR balanced sub-panel, 2023-01-01 to 2026-02-03, treated = CRZ; control = Manhattan above 60th St.
\end{table}

\begin{table}[H]\centering
\caption{Heterogeneity by hour and day type (ATR balanced sub-panel)}\label{tab:het}
\begin{tabular}{lrrrrr}\toprule
Subgroup & Coef & SE & 95\% CI low & 95\% CI high & p \\
\midrule
Peak hours (DiD x peak) & -0.4841 & 0.3390 & -1.1485 & 0.1803 & 0.1533 \\
Off-peak hours (DiD x off-peak) & -0.3137 & 0.5639 & -1.4189 & 0.7915 & 0.5779 \\
Weekday (DiD x weekday) & -0.4252 & 0.4101 & -1.2291 & 0.3786 & 0.2998 \\
Weekend (DiD x weekend) & -0.0000 & 0.0000 & -0.0000 & -0.0000 & 0.0017 \\
\bottomrule\end{tabular}
\flushleft\footnotesize\textit{Notes.} Outcome: $\log(1+\text{vol})$. Segment-clustered SE.
\end{table}

\begin{table}[H]\centering
    \caption{Pre-registered robustness battery}\label{tab:robust}
    \footnotesize
    \setlength{\tabcolsep}{3.5pt}
    \resizebox{\linewidth}{!}{%
    \begin{tabular}{@{}lrrrrrr@{}}\toprule
    Specification & Coef & SE & CI$_L$ & CI$_U$ & $p$ & N \\
    \midrule
    Headline (log vol; CRZ vs.\ M-above-60, weather) & -0.3927 & 0.3714 & -1.1206 & 0.3351 & 0.2902 & 2{,}520 \\
    Levels & -107.4211 & 61.8080 & -228.5647 & 13.7225 & 0.0822 & 2{,}520 \\
    asinh(vol) & -0.4306 & 0.4080 & -1.2303 & 0.3691 & 0.2912 & 2{,}520 \\
    Control& -0.3261 & 0.3331 & -0.9790 & 0.3268 & 0.3276 & 1{,}416 \\
    Drop bridges from treated set & -1.2399 & 0.0933 & -1.4229 & -1.0570 & 0.0000 & 2{,}112 \\
    Placebo treat.\ date 2024-01-05 & -1.0337 & 1.2045 & -3.3945 & 1.3271 & 0.3908 & 1{,}488 \\
    Anticipation cutoff 2024-11-14 & -0.3927 & 0.3714 & -1.1206 & 0.3351 & 0.2902 & 2{,}520 \\
    Weighted $1/n_{\mathrm{pre}}$ & -0.5198 & 0.4593 & -1.4201 & 0.3805 & 0.2578 & 2{,}520 \\
    \bottomrule\end{tabular}%
    }
    \flushleft\footnotesize\textit{Notes.} Each row re-estimates the headline TWFE-DiD specification on the indicated alternative sample/outcome. All specifications: segment FE, dow$\times$hour FE, weather and holiday controls; segment-clustered SE.
    \end{table}

\begin{table}[H]\centering
\caption{Within-unit pre/post change at potential spillover units}\label{tab:spillover}
\begin{tabular}{lrrrrr}\toprule
Unit & Segments & N & Pre/Post coef (log) & SE & p \\
\midrule
Bridges (any in panel) & 3 & 1,344 & 0.2266 & 0.1079 & 0.0356 \\
Manhattan above 60th & 3 & 1,944 & 0.0449 & 0.2279 & 0.8437 \\
\bottomrule\end{tabular}
\flushleft\footnotesize\textit{Notes.} Each row is a within-unit pre/post DiD: the coefficient on $\text{Post}_d$ within the indicated subgroup, with segment FE, dow$\times$hour FE, and weather. Positive coefficients indicate increased volume after Jan 5, 2025 -- consistent with route substitution.
\end{table}

\clearpage
\singlespacing
\bibliographystyle{apalike}
\bibliography{refs}

\end{document}